\documentclass[journal]{IEEEtran}
\usepackage[usenames]{color}
\usepackage{epsfig}
\usepackage{graphics}
\usepackage{amsmath}
\usepackage{amssymb}
\usepackage{multirow}
\usepackage{cite}
\usepackage{array}
\usepackage{pslatex} 
\usepackage{url}
\usepackage{caption}
\usepackage{lineno}
\usepackage{graphicx}  
\usepackage{setspace}
\usepackage{tikz}
\usepackage{hhline}
\usepackage{mathtools}
\usepackage[letterpaper]{geometry}

\usepackage{amsmath,amssymb,amsfonts}
\usepackage{algorithmic}
\usepackage{textcomp}
\usepackage{tabularx}

\usepackage{supertabular}
\usepackage{epstopdf}

 \newcommand{\flogo}{\includegraphics[height=18pt]{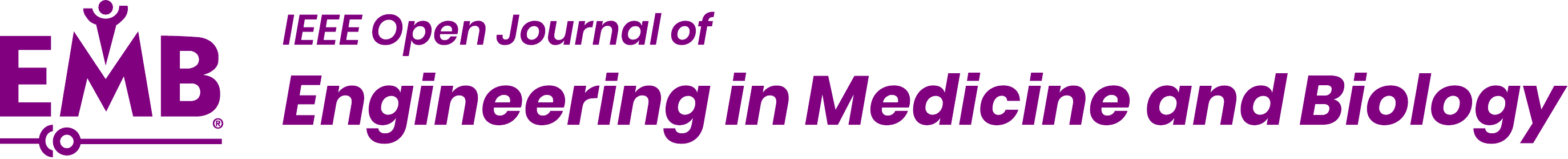}
 }
\ifCLASSINFOpdf
\else
\fi
\usepackage[english]{babel}
\usepackage[utf8]{inputenc}
\usepackage{fancyhdr}
\begin{document}
%

%
%
%

\title{\textcolor{violet}{Continuous Speech for Improved Learning Pathological Voice Disorders\vspace{0.25cm}}}
\author{Syu-Siang Wang, Chi-Te Wang, Chih-Chung Lai, \\ 
	Yu Tsao, \IEEEmembership{Senior Member, IEEE}, Shih-Hau Fang, \IEEEmembership{Senior Member, IEEE}
	\thanks{Chi-Te Wang is with the Department of Otolaryngology Head and Neck Surgery, Far Eastern Memorial Hospital, Taiwan. Yu Tsao is with the Research Center for Information Technology Innovation, Academia Sinica, Taiwan. Syu-Siang Wang, Chih-Chung Lai, Chi-Te Wang, and Shih-Hau Fang are with the Department of Electrical Engineering, Yuan Ze University, Taiwan (Email: shfang@saturn.yzu.edu.tw). 
		The study protocol was approved by the  Research Ethics Review Committee of Far Eatsern Memorial Hospital (FEMH-IRB No. 109063-E), date of approval: May 10th 2020.}}

\maketitle\thispagestyle{fancy}

\begin{abstract}
\textit{Goal:} Numerous studies had successfully differentiated normal and abnormal voice samples. Nevertheless, further classification had rarely been attempted. This study proposes a novel approach, using continuous Mandarin speech instead of a single vowel, to classify four common voice disorders (i.e. functional dysphonia, neoplasm, phonotrauma, and vocal palsy). \textit{Methods:} In the proposed framework, acoustic signals are transformed into mel-frequency \textcolor{black}{cepstral} coefficients, and a bi-directional long-short term memory network (BiLSTM) is adopted to model the sequential features. 
The experiments were conducted on a large-scale database, wherein 1,045 continuous speech were collected by the speech clinic of a hospital from 2012 to 2019. \textit{Results:} Experimental results demonstrated that the proposed framework yields significant accuracy and unweighted average recall improvements of 78.12–89.27\% and 50.92–80.68\%, respectively, compared with systems that use a single vowel. \textit{Conclusions}: The results are consistent with other machine learning algorithms, including gated recurrent units, random forest, deep neural networks, and LSTM.
The sensitivities for each disorder were also analyzed, and the model capabilities were visualized via principal component analysis. 
An alternative experiment based on a balanced dataset again confirms the advantages of using continuous speech for learning voice disorders.
\\
\par


\textit{Impact Statement—} Deep learning can detect common voice disorders using continuous speech. Future practice can screen patients who truly need hospital visits and reduce unnecessary medical demands, especially during COVID-19 pandemic.

\end{abstract}


%
\IEEEpeerreviewmaketitle


\section{INTRODUCTION}
\IEEEPARstart{V}oice \textcolor{black}{disorders are one of the most common health complaints, with the lifetime prevalence as high as 30\% in the general population \cite{roy2004prevalence, mehta2015using}.} Therefore, in recent decades, automatic detection of voice pathologies gathered much academic interest, and recent works verified the possibility by using the machine-learning-based classifiers, and acoustic signal features \textcolor{black}{\cite{arias2010automatic, 8337897,markaki2009using,hammami2016pathological,barche2020towards, barche2021comparative}}. 

For example, the works in \cite{ali2016voice} and \cite{muhammad2017voice} extracted vocal-fold-related acoustic features and combined them with a support vector machine for voice pathology detection.
The work in \cite{al2017investigation} performed correlation analyses on the sub-band signal to detect pathological voices. In addition,
deep learning and convolutional neural networks were investigated for pathological voice detection \cite{wu2018deep, mohammed2020voice, fang2018detection, wu2018convolutional, dahmani2018glottal, dahmani2017vocal, hemmerling2017voice}.
The recent works applied unsupervised domain adaptation to address the hardware variation \cite{hsu2018robustness}.
\textcolor{black}{Various acoustic features, including cepstral features \cite{pishgar2018pathological, pham2018diagnosing}, 
vocal jitter \cite{grzywalski2018parameterization},
and entropy \cite{al2017voice}} were also investigated in the literature. 
The IEEE Big Data conference held an international competition in Seattle 2018, called FEMH-Challenge, \textcolor{black}{in which voice pathology detection systems} from different research groups worldwide are evaluated empirically on the same dataset\textcolor{black}{, which was published} by Far Eastern Memorial Hospital (FEMH), Taiwan \cite{ramalingam2018ieee}. 
This competition established a systematic evaluation methodology with rigorous metrics 
for the comparison of voice disorders detection in fair conditions, and over one hundred teams participated in this challenge
\cite{bhat2018femh, degila2018ucd, arias2018byovoz, pham2018diagnosing, islam2018transfer, grzywalski2018parameterization, ramalingam2018ieee}.

Although numerous published studies had successfully differentiated normal and abnormal voice samples, further classification \textcolor{black}{has rarely been attempted.}
One possible reason may be the limitation of \textcolor{black}{the single vowel speech signal.} 
\textcolor{black}{Therefore, sustained vowels were investigated for
the voice disorder classification, and voice quality measures \cite{arias2019multimodal, mohammed2020voice, fujimura2019discrimination, cordeiro2015continuous,ali2016automatic,ali2016automatic}. }
The running speech from both English and Arabic databases was considered as well in \cite{mesallam2017development}, and the associated speech features were evaluated in terms of voice-disorder classifiers. 
Personal habits and behaviors, such as laryngeal tumors caused by long-term use of tobacco and alcohol, were studied in \cite{hashibe2009interaction}. 
The previous works combined acoustic signals and medical records
to classify three typical voice disorders, including glottic neoplasm, phonotraumatic lesions, and vocal paralysis \cite{tsui2018demographic, fang2019combining}. 
Although the results have confirmed the effectiveness of incorporating diverse information, 
using medical records or questionnaire data still requires an alternative human effort for data collection and 
a laptop or device for patients typing during the testing phase. These procedures may result in considerable extra costs for both the patient and society.
From a health science perspective, people should minimize the contact possibility, especially during the COVID-19 pandemic period.

In general, using the acoustic signal is believed to be the easiest and the most convenient way for the noninvasive screening of voice disorders. \textcolor{black}{Recent studies also demonstrated that reading a text passage can significantly reveal larger ranges of fundamental frequency and sound pressure level (i.e. intensity) \cite{chen2007sex,yen2021mandarin}.}
This motivates us to use continuous speech instead of a single vowel for this task because the multiple syllables may provide richer information to improve the performance.
On the other hand, some specialists, such as vocalists, could easily prevent vowel-based disorder recognition systems from working correctly by altering the way of vocalization. 
Fortunately, \textcolor{black}{altering abdominal vocalization for a long time is difficult} during the
Chinese syllable transitions, even for vocalist experts.
Thus, an alternative advantage of using continuous speech is that
it may provide valid detection information resistant to unintentionally abdominal/dantian vocalization from vocalist experts.

This study proposes a novel pathological voice classification approach using continuous speech.
To the best of our knowledge, this is the first study using sentence-based Mandarin speech signals to classify voice disorders automatically.
In the proposed framework, acoustic signals are transformed into temporal mel-frequency cepstral coefficients (MFCCs) and a bi-directional long-short term memory network (BiLSTM) is adopted to model the sequential features. 
The experiments were conducted on the large-scale Far Eastern Memorial Hospital (FEMH) voice disorder database, wherein all speech recordings were collected by the speech clinic of FEMH from 2012 to 2019.
\textcolor{black}{There is 1,045 continuous speech, each including 7 Chinese sentences, and 1,061 single vowel voice recordings,} distributed to functional dysphonia (FD), neoplasm, phonotrauma, and vocal palsy disorders. \textcolor{black}{These four types of vocal diseases are the most common diagnosis for patients with hoarseness \cite{stemple2018clinical}.} Notably, FD is subjective dysphonia but normal in endoscopic findings. 
For the classification task, experimental results demonstrated that the proposed framework yields significant accuracy and unweighted average recall (UAR) improvements of 78.12–89.27\% and 50.92–80.68\%, respectively, compared with systems that use only a single vowel. 
The results show that the dynamic models with memory architecture successfully extract the robust features from continuous speech, thus improving the effectiveness in the voice disorders classification task.

The results are consistent with other machine learning algorithms, including GRU (Gated Recurrent Unit), RF (Random Forest), DNN (deep neural networks), and LSTM.
The sensitivities for each disorder were presented and analyzed.
Results show that BiLSTM provides the highest sensitivity scores on FD (86.25\%) and vocal palsy (68.00\%), respectively, and competitive performance on neoplasm and phonotrauma.
The experiments carried out principal component analysis (PCA) to demonstrate the classified capability from testing a model on the continuous-speech corpus. 
Finally, the experiments were conducted on an alternative public FEMH-Challenge database, which is more balance in the categories with fewer samples.
The results again confirm the advantages of the proposed approach using continuous speech.

\section{MATERIALS AND METHODS }

\subsection{Database Description} 
We conducted our experiments on the FEMH voice disorder database, wherein all speech recordings were collected by the speech clinic of FEMH from 2012 to 2019. A seven-sentence script that was proposed in \cite{chen1996phonetograms} was applied to prepare the FEMH database. For each sentence, there are 1,045 voice-disorder recordings. In addition to these continuous speeches, 1,061 /$a$/-phone voices were also recorded for providing additional samples. The length of each /$a$/ sound was about three seconds. The distribution of these voice samples with respect to FD, neoplasm, phonotrauma, and vocal palsy was listed in Table \ref{tab:FEMH_data}. 
\textcolor{black}{All waveforms were recorded at 44,100 Hz sampling rate with a 16-bit integer resolution and using a high-quality microphone (model: SM58, Shure, IL) with a digital amplifier (MDVP, Model 4500, Kay Elemetrics) under a background noise level between 40 and 45 dBA.} For each of seven-sentence-speech- and /$a$/-sound-corpus, 80\% voices were applied to form the training and developing set while the remaining 20\% sounds were used to provide the testing set. Notably, there is no overlapped speaker between training and testing sets. 

\begin{table}[!ht]
	\addtolength{\tabcolsep}{3pt}
	\centering
	\caption{The number of voice-disorder samples for the FEMH database.}
	\label{tab:FEMH_data}
	\begin{tabularx}{\columnwidth}{>{\centering}m{2cm}|>{\centering}m{1.3cm}>{\centering}m{1.3cm}>{\centering}m{1.3cm}>{\centering\arraybackslash}X}
		\hline
		\hline
		& \textbf{\scriptsize FD} & \textbf{\scriptsize Neoplasm} & \textbf{\scriptsize Phonotrauma} & \textbf{\scriptsize Vocal palsy}\\
		\hline
		\textbf{\scriptsize continuous speech} & 100 & 103 & 718 & 124 \\
		\textbf{/$a$/ sound} & 100 & 102 & 735 & 124\\
		\hline
		\hline
	\end{tabularx}
\end{table}

\textcolor{black}{On the other hand, an alternative dataset published for international competition (called FEMH challenge)} is employed for further evaluation in the experiments. 
The statistics for the FEMH-Challenge database are organized in Table \ref{tab:femhchallenge}. 
Comparing Table \ref{tab:femhchallenge} with Table \ref{tab:FEMH_data}, it can be observed that this database is relatively small but more balance among the three categories.
It allows us to provide a fair comparison with existing methods from different perspectives. 
To be more specific, this database is composed of 150 /$a$/-vowel sounds that were pronounced by 150 different patients and classified into neoplasm, phonotrauma, and vocal palsy. In the meanwhile, these speakers were asked to utter the seven-sentence script.
The front-end data collection procedure, environment, and separation were identical in these two datasets. 

\begin{table}[!ht]
	\centering
	\caption{The number of voice-disorder samples for the FEMH-Challenge database.}
	\label{tab:femhchallenge}
	\begin{tabularx}{\columnwidth}{>{\centering}m{2.5cm}|>{\centering}m{2cm}>{\centering}m{2cm}>{\centering\arraybackslash}X}
		\hline
		\hline
		& \textbf{\scriptsize Neoplasm} & \textbf{\scriptsize Phonotrauma} & \textbf{\scriptsize Vocal palsy}\\
		\hline
		\textbf{\scriptsize continuous speech} & 40 & 60 & 50 \\
		\textbf{/$a$/ sound} & 40 & 60 & 50 \\
		\hline
		\hline
	\end{tabularx}
\end{table}

\begin{figure*}[!t]
	\centering 
	\centerline{
		\includegraphics[scale=1]{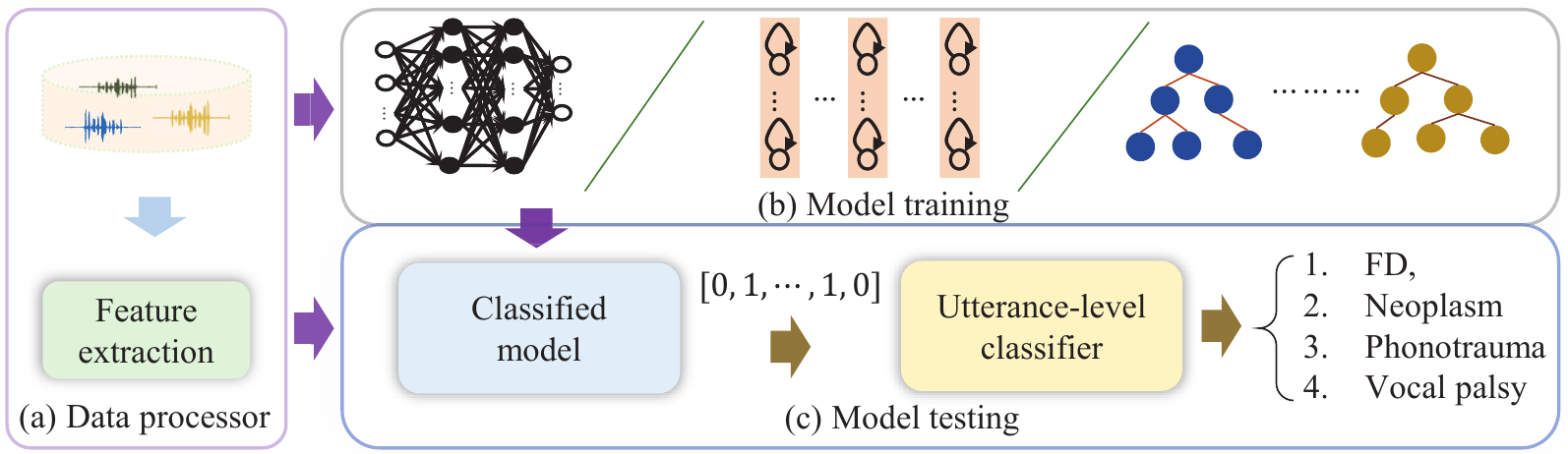}}
	\caption{The block diagram of the proposed system, which comprises of three stages: (a) data processing, (b) model training, and (c) testing stages.} 
	\label{fig:blockdiagram}
\end{figure*}

\subsection{Proposed method}
Figure \ref{fig:blockdiagram} shows the block diagram of the proposed approach, including data processing, model training, and online testing stages. 
From this figure, an acoustic feature extraction module is applied to extract speech features from the input waveform. 
The supervised model is performed by mapping the relationship between the feature and frame-level classification label in the following training stage. The prediction is then achieved by passing the speech feature through the model followed by the utterance-level classification in the final testing stage.
\textcolor{black}{The Research Ethics Review Committee of Far Eastern Memorial Hospital approved the study protocol (FEMH-IRB No. 109063-E), date of approval: May 10th, 2020.
The details of these stages are described in the following subsections.}

\subsection{Data Processing Stage}
In the data processing stage, voice activity detection and feature extraction operations were utilized for pre-processing the input acoustic waveform.

\subsubsection{Voice Activity Detection}
Voice segments of an utterance were detected by the voice activity detection (VAD) technique, which assumed uncorrelated noises degraded the speech. The Gaussian statistical model was used to model each voice and non-voice component for calculating the likelihood ratio. Those speech present frames were determined thereafter with respect to the predefined threshold. In this study, the threshold was set to 0.9, and the VAD we applied was developed from the VOISEBOX tool\footnote{http://www.ee.ic.ac.uk/hp/staff/dmb/voicebox/voicebox.html}. In addition, the prior and posterior signal-to-noise ratio (SNR) for VAD was calculated through the minimum-mean square error short-time spectral amplitude estimator method.

\subsubsection{Feature Extraction}
MFCCs were extracted from those speech segments in the following steps. The hamming-window framing function was applied to split input voice segments into a sequence of frame signals, wherein the frame length and hop size were 32ms and 16ms, respectively. A high-pass filter was applied for each frame to pre-emphasize the input and then decomposed to magnitude spectra through the discrete Fourier transformation. The spectra were filtered through a set of mel filters and then processed by logarithm and power operations to generate 26 filter-bank coefficients. A 13-dimensional static MFCC of a frame was then made-up from this 26-dimensional vector by applying discrete cosine transformation. The MFCC-delta was derived from static MFCC in terms of the delta operator. Finally, the proposed algorithm used 26-dimensional MFCC (MFCC and MFCC-delta) for this task.

\subsection{Model Training Stage}\label{sec:mdl}
\begin{figure}[!t]
	\centering 
	\centerline{
		\includegraphics[scale=1]{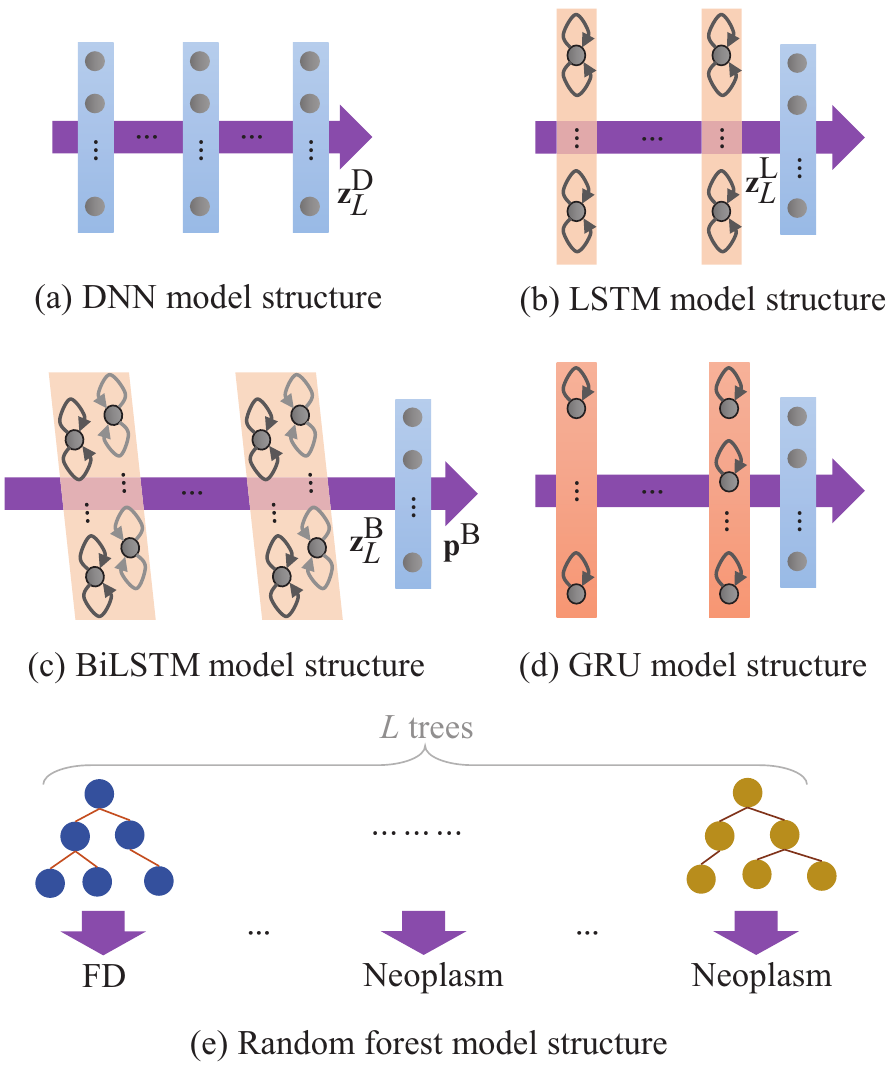}}
	\caption{\textcolor{black}{The (a) DNN, (b) LSTM, (c) BiLSTM, (d) GRU, and (e) RF model structure.}}\label{fig:mdlsru}
\end{figure}

The goal of the paper is to identify the pathological voice from an input continuous speech. Specifically, the model input is an MFCC frame, while the output is the frame-wise one-hot label vector. Because the varied length of different recordings results in altered MFCC size, we expect models to achieve two characters: (1) effectively extracted representative features from the size-varied input and (2) handling the contextual information from the input waveform for classifying the voice. 
Thus, we investigated five learning-based models in the proposed framework\textcolor{black}{, where the model structure is depicted in Fig. \ref{fig:mdlsru}, and introduced in the following subsections.}

\subsubsection{Deep Neural Network}
A DNN model \textcolor{black}{illustrated in Fig. \ref{fig:mdlsru} (a)} comprising of multiple hidden layers was leveraged to convert a 26-dimensional MFCC frame to the frame-level one-hot label vector. The output of each hidden layer was modified by using the relu activation function, $\sigma$, while the output value of DNN was scaled to the range between 0 and 1 by the softmax function. For an arbitrary $\ell$th hidden layer, the formulation of input-output ($\mathbf{z}_{\ell}^\mathfrak{D}$, $\mathbf{z}_{\ell-1}^\mathfrak{D}$) is showed by
\begin{equation}
	\mathbf{z}_{\ell}^\mathfrak{D}=\sigma_\ell\{h_{\ell}(\mathbf{z}_{\ell-1}^\mathfrak{D})\},
\end{equation}
where $h_\ell(\cdot)$ is the linear transformation function. The model parameters were then derived in terms of the cross-entropy cost function in the training process. Because the non-linear transformation capability of a DNN, we believe that the model can be utilized to extract the representative acoustic features from vowel voice and the continuous speech in this classification task.

\subsubsection{Long Short-Term Memory}
The LSTM-classification system \textcolor{black}{depicted in Fig. \ref{fig:mdlsru} (b)} comprised $L$ LSTM and a feed-forward hidden layers. By passing the $n$th MFCC frame across all LSTM hidden layers, the contextual feature $\mathbf{z}_L^{\mathfrak{L}}[n]$ in the output side of LSTM were derived by jointly considering the previous $n$ MFCC input vectors. $\mathbf{z}_L^{\mathfrak{L}}[n]$ was then placed on the input of a feed-forward layer and then provided the predicted likelihood at the output of the system. For performing an LSTM-classification system, the cross-entropy cost function was implemented to calculate the distance between ground truth label vectors and model predictions and minimized thereafter to provide each LSTM and feed-forward layer parameters for the next testing stage.

\subsubsection{Bi-directional Long Short-term Memory}
With respect to an input MFCC feature, a BiLSTM and feed forward blocks were carried out for performing the BiLSTM-classified system\textcolor{black}{, which was showed in Fig. \ref{fig:mdlsru} (c). From the figure, the} $\ell$th hidden layer input-output relationship ($\mathbf{z}_\ell^{\mathfrak{B}}[n]$, $\mathbf{z}_{\ell+1}^{\mathfrak{B}}[n]$) at the $n$th frame in the BiLSTM block is formulated as
\begin{equation}\label{eq:bilstm}
	\begin{aligned}
		\mathbf{z}_{\ell}^{\mathfrak{b}}[n]&=\text{LSTM}^{\mathfrak{b}}_{\ell}\{\mathbf{z}_{\ell}^{\mathfrak{B}}[n]\},\\
		\mathbf{z}_{\ell}^{\mathfrak{e}}[n]&=\text{LSTM}^{\mathfrak{e}}_{\ell}\{\mathbf{z}_{\ell}^{\mathfrak{B}}[n]\},\\
		\mathbf{z}_{\ell+1}^{\mathfrak{B}}[n]&=\mathbf{z}_{\ell}^{\mathfrak{b}}[n]+\mathbf{z}_{\ell}^{\mathfrak{e}}[n],
	\end{aligned} \:\:\ell=1,\cdots,L,
\end{equation}
where LSTM$^{\mathfrak{b},\mathfrak{e}}\{\cdot\}$ represents an LSTM cell. From Eq. \eqref{eq:bilstm}, not only extracting features from the beginning of an MFCC (i.e. LSTM$^{\mathfrak{b}}_{\ell}\{\cdot\}$), the BLSTM block performed acoustic representations by further considering the speech structure from the very end of an utterance (i.e. $LSTM^{\mathfrak{e}}_{\ell}\{\cdot\}$). Therefore, contextual structures from both ends of MFCC were modeled at the $n$th output $\mathbf{z}_{\ell+1}^{\mathfrak{B}}[n]$ in the BiLSTM block for the following feed forward layers.

The one-layer feed forward block was used in this study for performing the output prediction $\mathbf{p}^{\mathfrak{B}}[n]$ with respect to $\mathbf{z}_{L+1}^{\mathfrak{B}}[n]$ and is formulated by 
\begin{equation}\label{eq:bilstmff}
	\mathbf{p}^{\mathfrak{B}}[n]=\text{softmax}\{\mathbf{W}\mathbf{z}_{L+1}^{\mathfrak{B}}[n]+\mathbf{b}\},
\end{equation}
where $\mathbf{W}$ is the weight matrix while $\mathbf{b}$ represents the bias vector. 
We then derived both BiLSTM and feed-forward blocks by leveraging 
the softmax activation function and the cross-entropy cost function.

\subsubsection{Gated Recurrent Unit}
GRU proposed in \cite{cho2014properties} was explored in this study to make the recurrent unit capture the long time dependency and extract the robust features from MFCC. Similar to LSTM, the GRU model \textcolor{black}{that was illustrated in Fig. \ref{fig:mdlsru} (d)} was performed in terms of the gated mechanism to control the information flow through the model. However, fewer gates were used in GRU than those in LSTM, and thereby reducing the size of the model as well as the system latency on processing input data \cite{ravanelli2018light}. 
GRU has been vastly investigated on speech recognition systems. For example, the work in \cite{tang2017memory} studied the behavior of LSTM and GRU and observed the robustness of GRU in noisy environments. The works in \cite{cheng2019automatic} and \cite{cheng2018output} proposed output-gate projected GRU (OPGRU) that applied an additional transformation matrix on the output GRU. According to these studies, the extracted OPGRU features further improved the recognized accuracy of an acoustic model while improving the decoding speed from the input waveform to the output phone-level sequence. 

Inspired by those works on speech recognition that recognized words from input speech, the GRU model was leveraged in this study for identifying the category of disease on the vocal tract through voice. Herein, we followed the work of \cite{cho2014properties} for constructing the GRU model, which contains $L$ hidden layers and a feed-forward layer followed by the softmax activation function. The system output is the frame-level predicted vectors. The cross-entropy cost function was applied for performing the model training procedure.

\subsubsection{Random Forest}
RF was composed of a series of decision trees \textcolor{black}{as showed in Fig. \ref{fig:mdlsru} (e),} wherein each tree created a decision rule for the classification. The rule was derived in terms of the training data attributed, whereas the number of leaf nodes for each tree algorithm was determined subsequently. A simple majority vote was then performed across all trees to shrink the predicting variance. In the training stage, $m$ random samples selected from training data were applied to achieve one unpruned classification and regression tree and split afterward at each node to improve classification accuracy. After performing all $T$ trees, an optimized fitting function was applied in the testing stage.

\subsection{Online Testing Stage}
In the online testing stage, an MFCC frame was put on the input side of a classification system that generated the predicted vector at the output terminal. The frame-level classification was then derived from the value in the vector that provided the highest class likelihood. All predictions were aggregated in the label predictor in Fig. \ref{fig:blockdiagram} and determined the final disease category by applying a majority vote.

\subsection{Model Training Setup}
Five learning models were investigated for the pathological-voice classification task in this study. \textcolor{black}{In addition, the five-fold learning strategy was applied to evaluate each machine-learning model.} For DNN, there are three hidden layers, and the size of each layer is 200. For LSTM, two hidden layers \textcolor{black}{with the dropout rate of 0.2 and 50 cells per layer were used for constructing the LSTM-classification system. 
Similar settings for performing LSTM model structure were also employed for providing GRU, two 50-cell GRU hidden layers with a 0.2 dropout rate.}

Meanwhile, two BiLSTM hidden blocks were used, where each block contains two 50-cell LSTM architectures. The dropout operation was also implemented for each of both LSTMs with a rate of 0.2. All deep-learning models were optimized by using the Adam optimizer with the learning rate of 0.001 for DNN and GRU while that of 0.0005 was used for LSTM and BiLSTM to achieve the optimized performance in the testing stage. As for the RF machine learning model, the bootstrap method is used to sample randomly from training data. In addition, twenty-six trees $T=26$ were used in RF for optimizing the classified results.

\subsection{Evaluation Metrics}
We evaluated the classified performance of the proposed system in terms of the overall accuracy, sensitivity, and unweighted average recall (UAR) metrics. Among these indexes, the score was calculated from the combination of true positive (TP), true negative (TN), false positive (FP), and false negative (FN). The accuracy score provided in Eq. \eqref{eq:acc} was performed to demonstrate the predicted correctness between prediction and truth. 
\begin{equation}
	\label{eq:acc}
	\text{Accuracy} = 100\%\times \frac{\text{TN}+\text{TP}}{\text{TN} + \text{TP} + \text{FN} + \text{FP}}
\end{equation}
The sensitivity value performed in Eq. \eqref{eq:Sn} was calculated for each of four voice-disorder classes ($\mathfrak{D}\in\{\mbox{FD},\mbox{ neoplasm},\mbox{ phonotrauma},\mbox{ vocal palsy}\}$). 
\begin{equation}
	\label{eq:Sn}
	\mbox{Sensitivity}_{\mathfrak{D}} = 100\%\times \frac{\mbox{TP}_{\mathfrak{D}}}{\mbox{TP}_{\mathfrak{D}}+\mbox{FN}_{\mathfrak{D}}}.
\end{equation}
Then, the final UAR score was obtained by averaging these sensitivity values and is showed in the following equation.
\begin{equation}
	\label{eq:UAR}
	\mbox{UAR} = 100\%\times\frac{\sum_{\mathfrak{D}}\mbox{Sensitivity}_{\mathfrak{D}}}{K},
\end{equation}
where $K=4$ was used for FEMH while $K=3$ was set for FEMH-Challenge.

\section{Results}

\subsection{Sentence Selection from FEMH}

The experiments first evaluate the effectiveness of each sentence from the classified performance.
In this experiment, we performed a BiLSTM-classified system on the sentence-based FEMH, that is 836 speech-label training pairs, and tested it on the associated testing data.
All Chinese characters of each sentence were transferred to Pinyin with tone marks, as shown in 
Table \ref{tab:sentence1}. 
From this table, the second sentence, ``lan3 lan3 de5 shuo1 le5 yi4 sheng1 ching3 jin4 lai2'', achieves the highest accuracy and UAR scores among seven Chinese sentences. 
The best score implies that this sentence provides the most balanced pronunciation effort, the articulate structure, and the classified information for both patients and systems. 
Therefore, this study used the second sentence as the continuous speech for the following evaluation subsections.

\begin{table}[!ht]
	\addtolength{\tabcolsep}{0.5pt}
	\centering
	\caption{Accuracy and UAR (\%) from each sentence based on the BiLSTM disease-classifier system. }
	\label{tab:sentence1}
	\begin{scriptsize}
		\scalebox{0.92}{
			\begin{tabular}{lcc}
				\hline
				\hline
				{\centering\textbf{Chinese sentence}}& \textbf{Accuracy} & \textbf{UAR}\\ \hline
				\textbf{\scriptsize wo3 ting1 dao4 you3 ren2 chiao1 men2}& 82.88& 75.39 \\ 
				\textbf{\scriptsize lan3 lan3 de5 shuo1 le5 yi4 sheng1 ching3 jin4 lai2}& \textbf{89.27}& \textbf{80.68} \\ 
				\textbf{\scriptsize men2 kai1 le5 wo3 kan4 dao4 yi4 ge5 nian2 ching1 ren2}& 84.63& 74.09 \\ 
				\textbf{\scriptsize shou4 chang2 de5 shen1 ti3 ming2 liang4 de5 yan3 jing1}& 86.23& 75.55 \\ 
				\textbf{\scriptsize hai2 you3 yi4 jhang1 cheng2 ken3 de5 lian3}& 80.17& 74.78 \\ 
				\textbf{\scriptsize kan4 ta1 lian3 shang4 de5 biao3 ching2 yi3 ji2 yan2 su4 de5 tai4 du4}& 82.13& 67.47 \\ 
				\textbf{\scriptsize jhen1 siang4 you3 shen2 me5 shih4 ching2 yao4 wo3 bang1 jhu4} & 81.37& 73.75 \\ 
				\hline
				\hline
		\end{tabular}}
	\end{scriptsize}
\end{table}

\begin{figure}[!ht]
	\centering 
	\centerline{
		\includegraphics[width=\columnwidth]{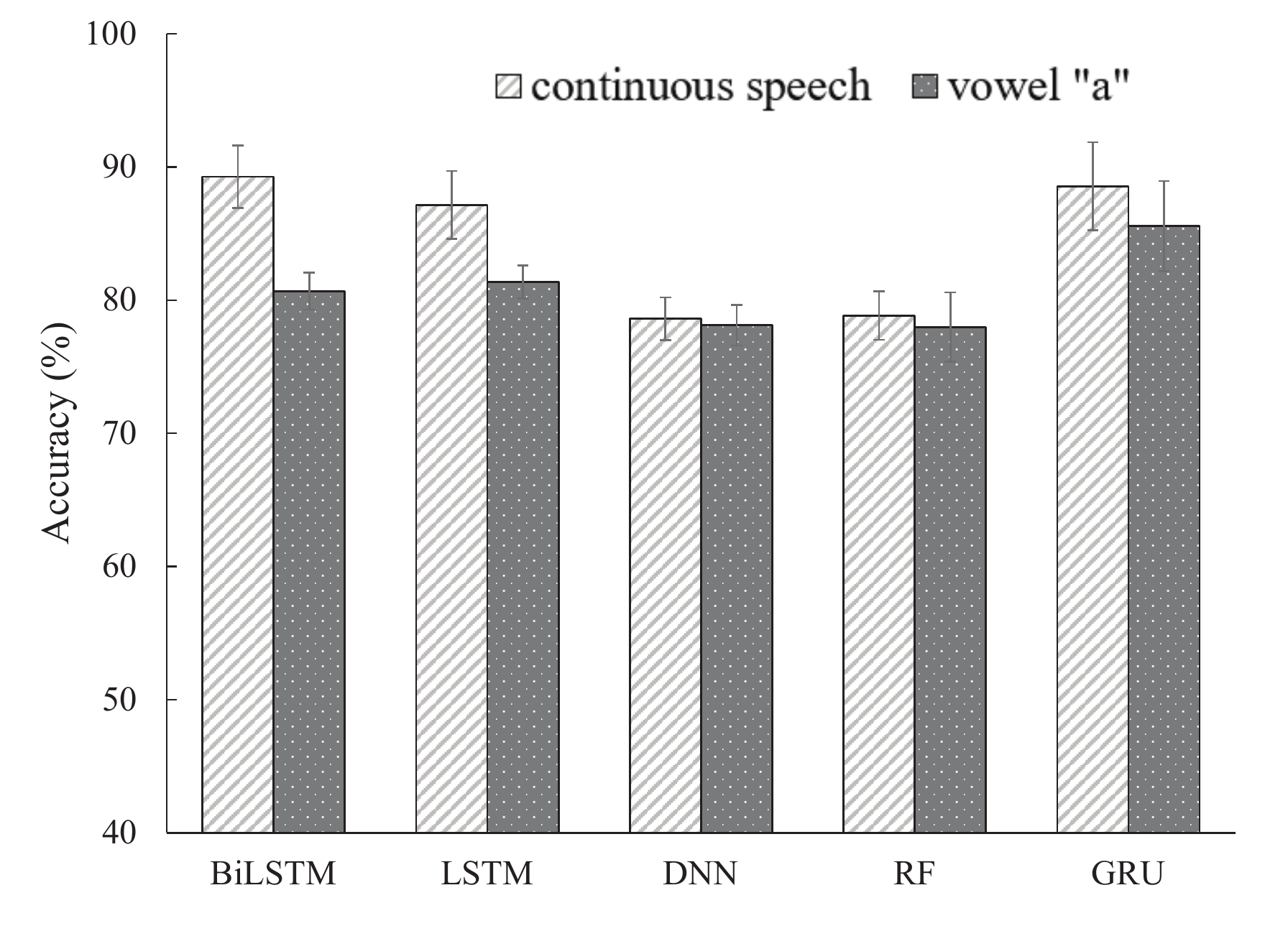}}
	\caption{Accuracy comparison between continuous-speech and /$a$/-phone using FEMH dataset with five machine learning algorithms}
	\label{fig:acc_all}
\end{figure}
\begin{figure}[!ht]
	\centering 
	\centerline{
		\includegraphics[width=\columnwidth]{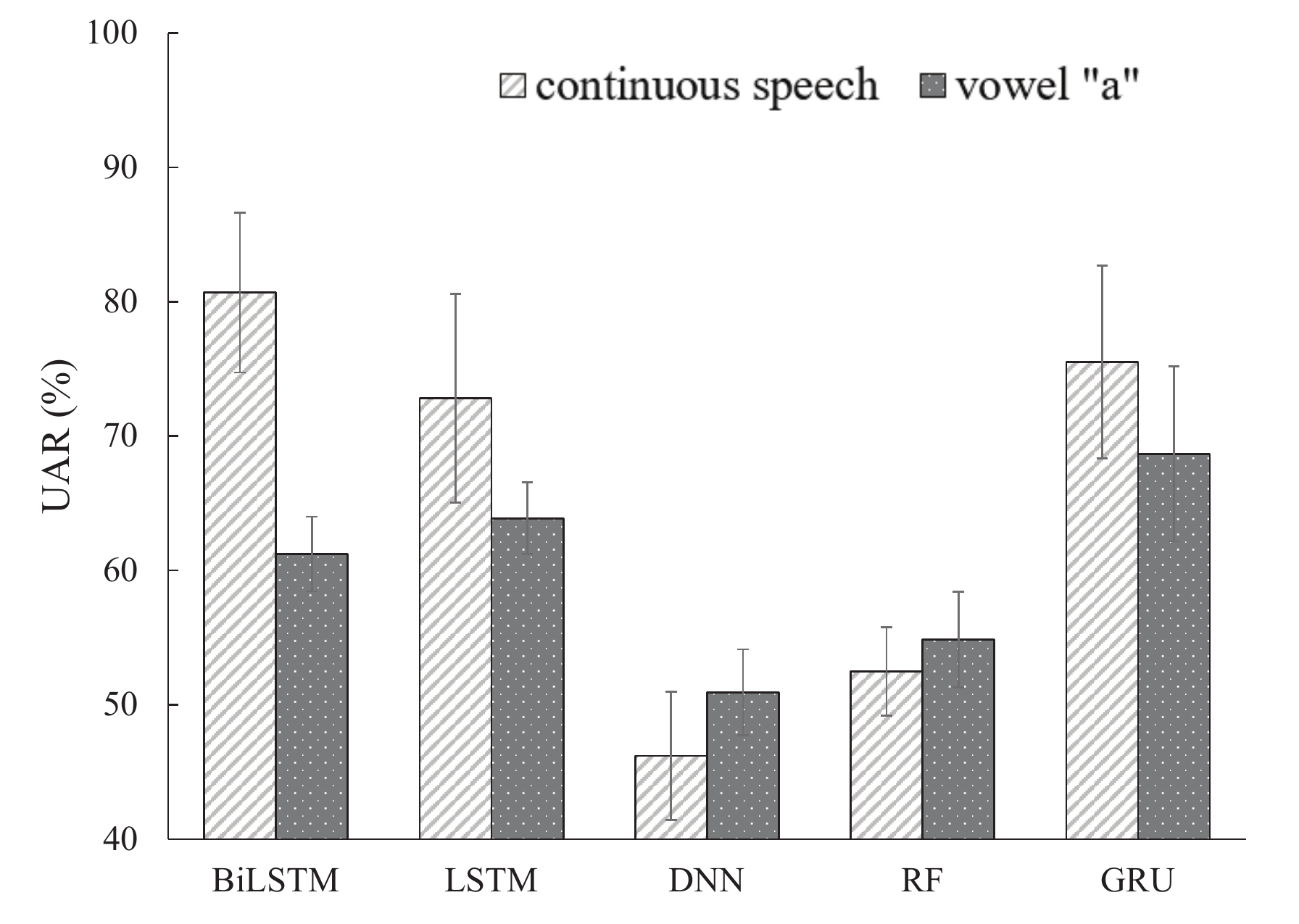}}
	\caption{UAR comparison between continuous-speech and /$a$/-phone using FEMH dataset with five machine learning algorithms} 
	\label{fig:uar_all}
\end{figure}

\subsection{Evaluations on FEMH Voice Disorder Database}
In this subsection, five machine learning classifiers (BiLSTM, GRU, LSTM, DNN, and RF systems) \textcolor{black}{introduced in Section \ref{sec:mdl}} were implemented using FEMH continuous-speech and /$a$/-vowel training corpus to classify four types of vocal disorders. 
Figures \ref{fig:acc_all} and \ref{fig:uar_all} illustrate the averaged accuracy scores and UAR, respectively, under continuous-speech and /$a$/-vowel testing conditions. 
Both figures show that continuous speech significantly outperforms the single /$a$/-vowel, 
except for the UAR using DNN and RF systems. 
The observation suggests that abundantly fine-structure and vowel-transition speech attributes in the continuous-speech further promote the model capability from identifying the pathological voice in the classification task. 
Because DNN and RF are static model structures, extracting information from the continuous speech is difficult. This may explain the exception in Fig. \ref{fig:uar_all}.
On the other hand, for those evaluations on the continuous-speech database, the classified scores from BiLSTM, GRU, and LSTM systems were better than those from DNN and RF. 
Three dynamic models perform similarly in the accuracy metrics while BiLSTM achieves the highest UAR.
Specifically, BiLSTM achieved the highest accuracy (89.27\%) and UAR (80.68\%). 
The results confirm that the dynamic models with memory architecture successfully extract the robust features from continuous speech, thus improving the effectiveness in the voice disorders classification task.

\begin{figure}[!ht]
	\centering 
	\centerline{
		\includegraphics[width=\columnwidth]{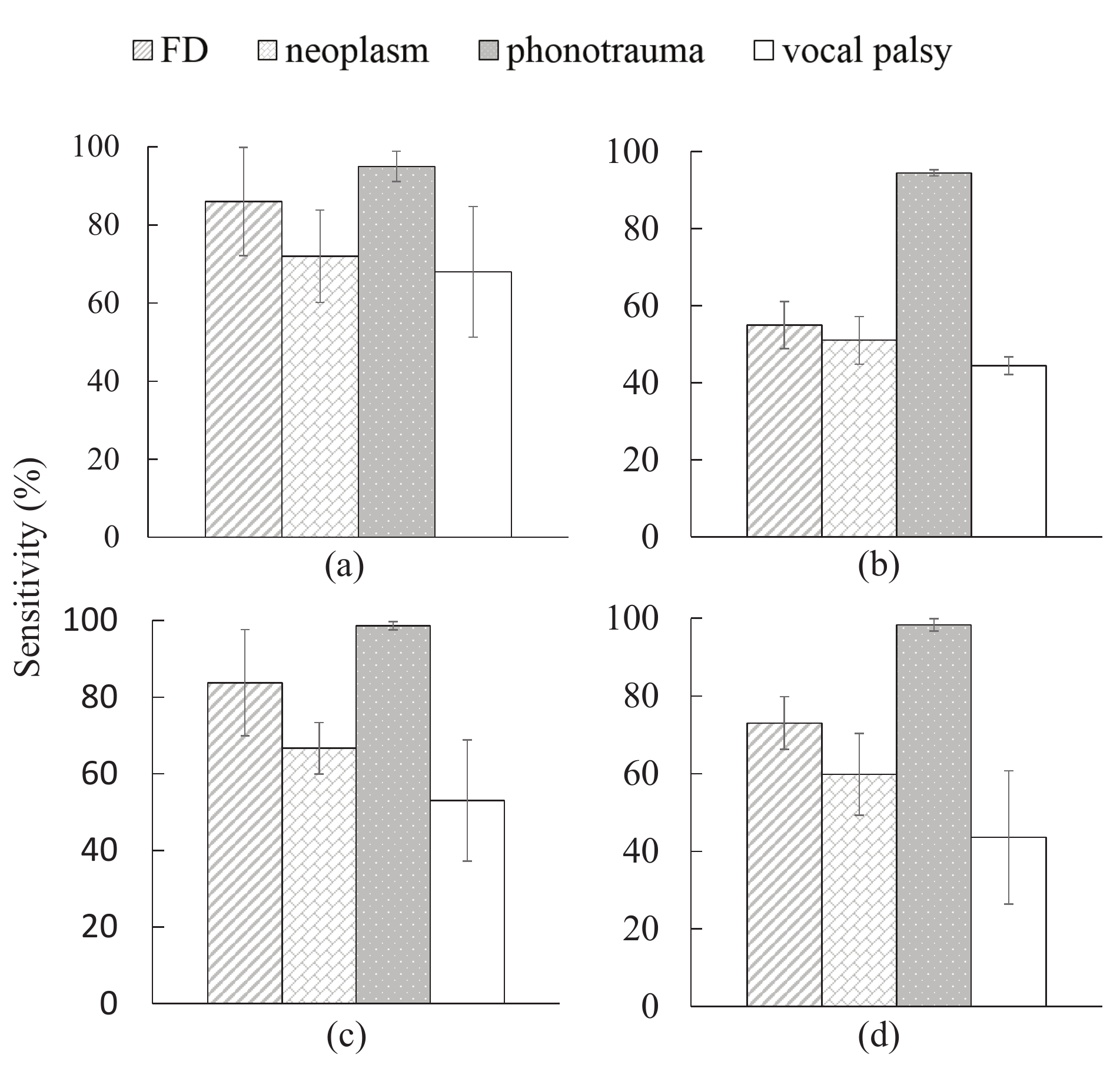}}
	\caption{Sensitivity for each vocal disorders (a) BiLSTM with continuous-speech (b) BiLSTM with single vowel (c) GRU with continuous-speech and (d) GRU with single vowel.} 
	\label{fig:sen_all}
\end{figure}

In addition to accuracy and UAR, the sensitivities for each disorder, including FD, neoplasm, phonotrauma, and vocal palsy, were presented and analyzed.
Figures \ref{fig:sen_all} (a) and (b) show the sensitivity of continuous-speech and single vowel using BiLSTM models while figures \ref{fig:sen_all} (c) and (d) report that using the GRU models.
In the viewpoint of disorder types, when comparing Figs. \ref{fig:sen_all} (a) with (b) and (c) with (d), both BiLSTM and GRU systems on continuous-speech corpus provided better classified results than those on /$a$/-vowel database. 
Again, Fig. \ref{fig:sen_all} show that both BiLSTM and GRU systems using continuous-speech provided better classified results than those on /$a$/-vowel ones. 
Comparing Figs. \ref{fig:sen_all} (a) with (c), BiLSTM provides the highest Sensitivity scores on FD (86.25\%) and vocal palsy (68.00\%), respectively, and competitive performance on neoplasm and phonotrauma. Comparing Figs. \ref{fig:sen_all} (a)(c) with (b)(d), only phonotrauma shows the comparable performance between continuous-speech and single vowel. The other three disorders were difficult to classify without continuous speech information. 
The result again confirms the effectiveness of applying continuous-speech corpus to classify voice disorders and the ability of BiLSTM to achieve the highest sensitivity. 
We also noted that scores on phonotrauma are higher than other types of diseases in all subfigures. One possible inference is that the extensive training samples for phonotrauma may result in the model bias issue. Fortunately, the BiLSTM model can effectively shrink the bias phenomena from the extracted speech features in this task.


\subsection{Evaluations on FEMH-Challenge Database}

\begin{figure}[!t]
	\centering 
	\centerline{
		\includegraphics[width=\columnwidth]{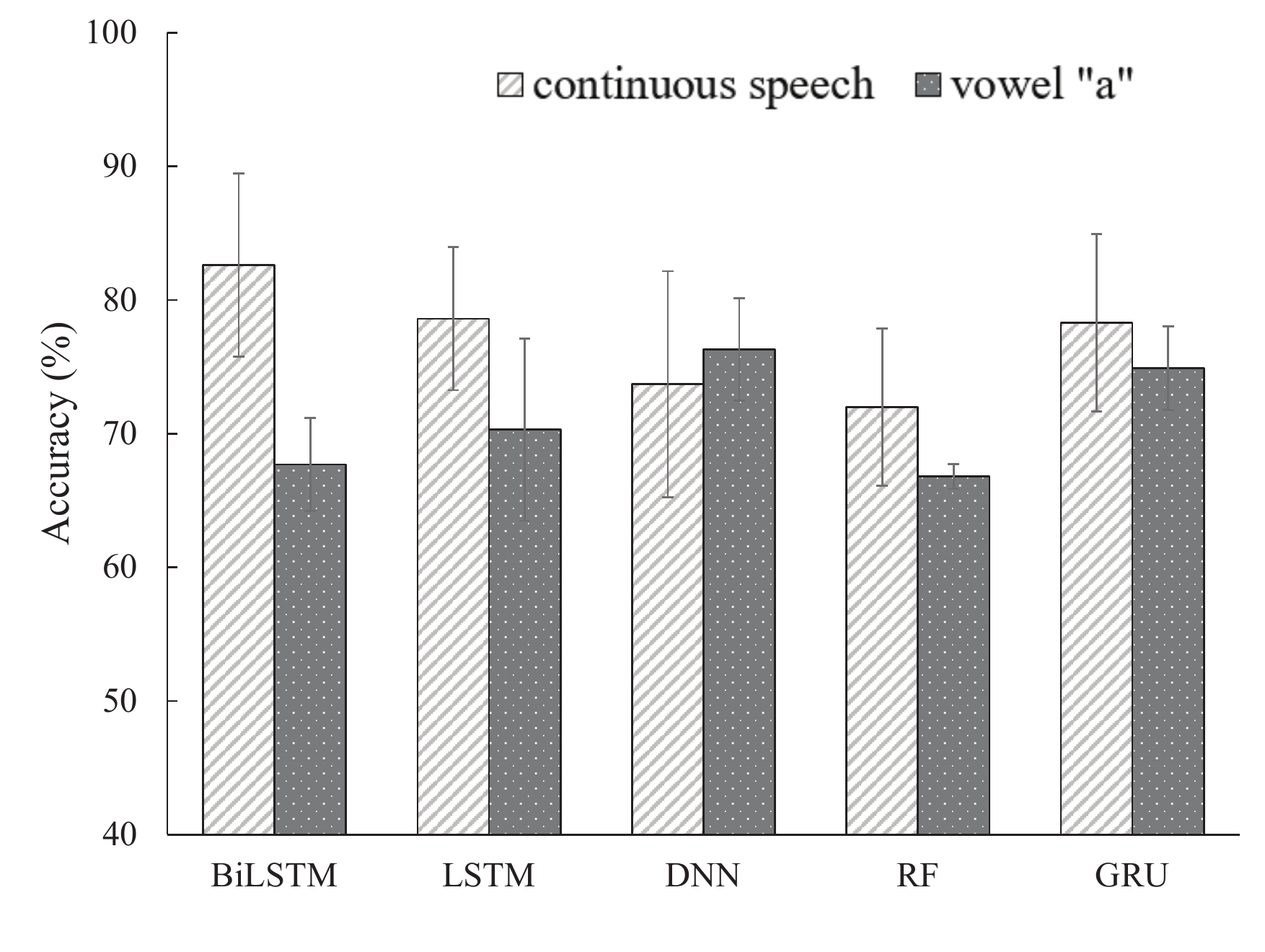}} 
	\caption{Accuracy comparison between continuous-speech and /$a$/-phone using FEMH-Challenge dataset with five machine learning algorithms} 
	\label{fig:2018acc}
\end{figure}

\begin{figure}[!t]
	\centering 
	\centerline{
		\includegraphics[width=\columnwidth]{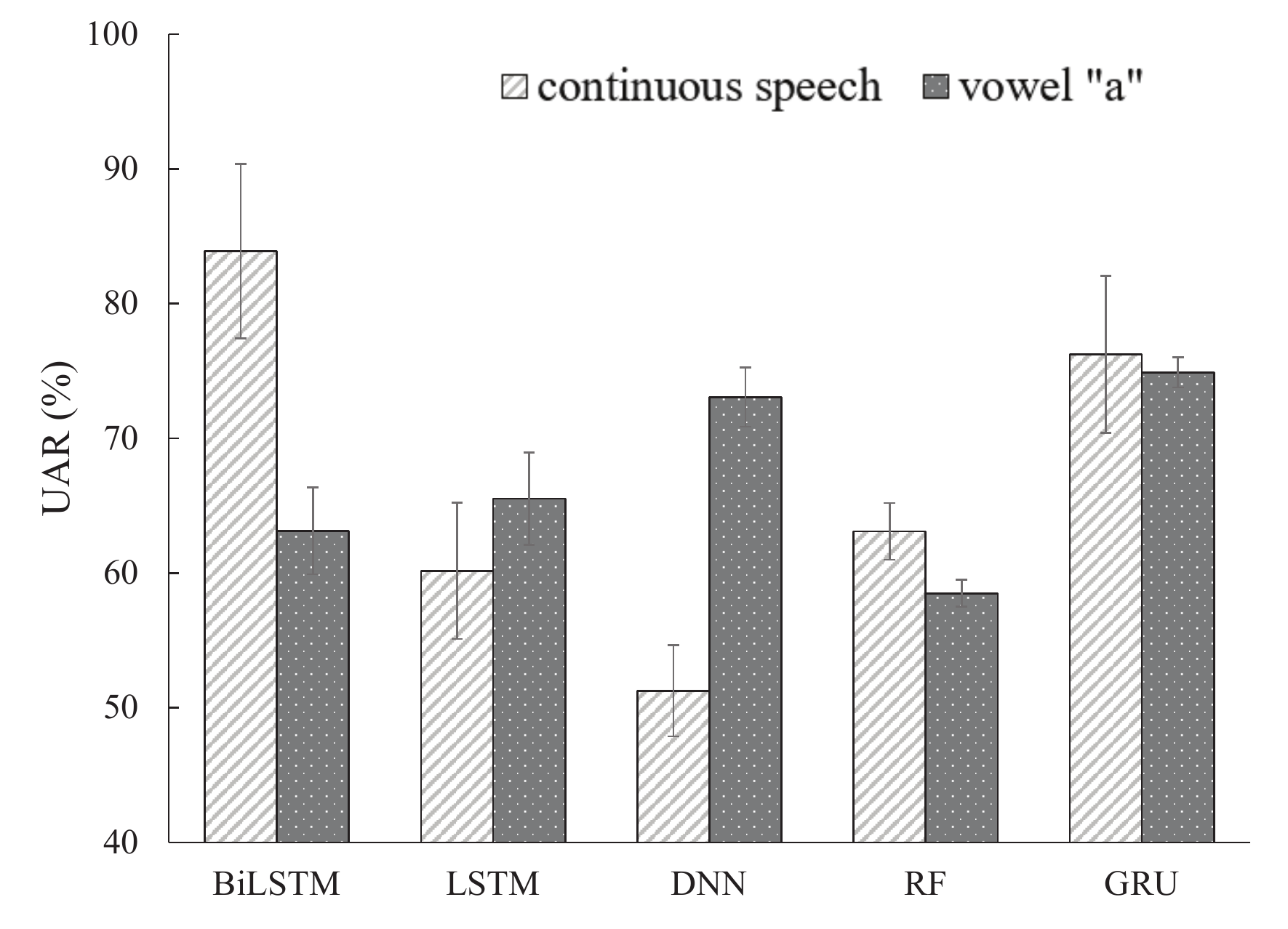}} 
	\caption{UAR comparison between continuous-speech and /$a$/-phone using FEMH-Challenge dataset with five machine learning algorithms} 
	\label{fig:2018uar}
\end{figure}

In this subsection, a balanced FEMH-Challenge database comprising neoplasm, phonotrauma, and vocal palsy voice samples was applied for performance evaluation. 
Tables \ref{tab:femhchallenge} and \ref{tab:FEMH_data} showed the different samples sizes
of FEMH-challenge with the full FEMH dataset. 
Figures \ref{fig:2018acc} and \ref{fig:2018uar} depicted the accuracy and UAR scores of continuous-speech and /$a$/-vowel sound using this balance dataset. 
Experimental results demonstrated that the proposed framework yields significant accuracy improvements compared with systems that use only a single vowel; the only exception is DNN. 
The results are consistent with the previous section.
Both experiments from a large-scale FEMH or a balance FEMH-Challenge dataset confirm the advantage of the proposed approach using continuous speech.
However, Fig. \ref{fig:2018uar} shows that only BiLSTM provides the comparable improvement of UAR as that of accuracy, and the DNN-based approach does not work well for this performance metric.
The results imply that the DNN-classified system couldn't effectively leverage the input's contextual information to provide decent output predictions. 
Meanwhile, the decreased performance of LSTM and DNN may reflect the impact of small dataset size. The training samples are insufficient to train a relatively complex memory structure in LSTM.
On the other hand, GRU remains a similar performance because it reduces the number of gates in a neural network.  
The results also demonstrate that BiLSTM performs well on a small database, achieving the highest accuracy and UAR in both experiments.

\section{Discussion}

\begin{figure}[!t]
	\centering 
	\centerline{
		\includegraphics[width=\columnwidth]{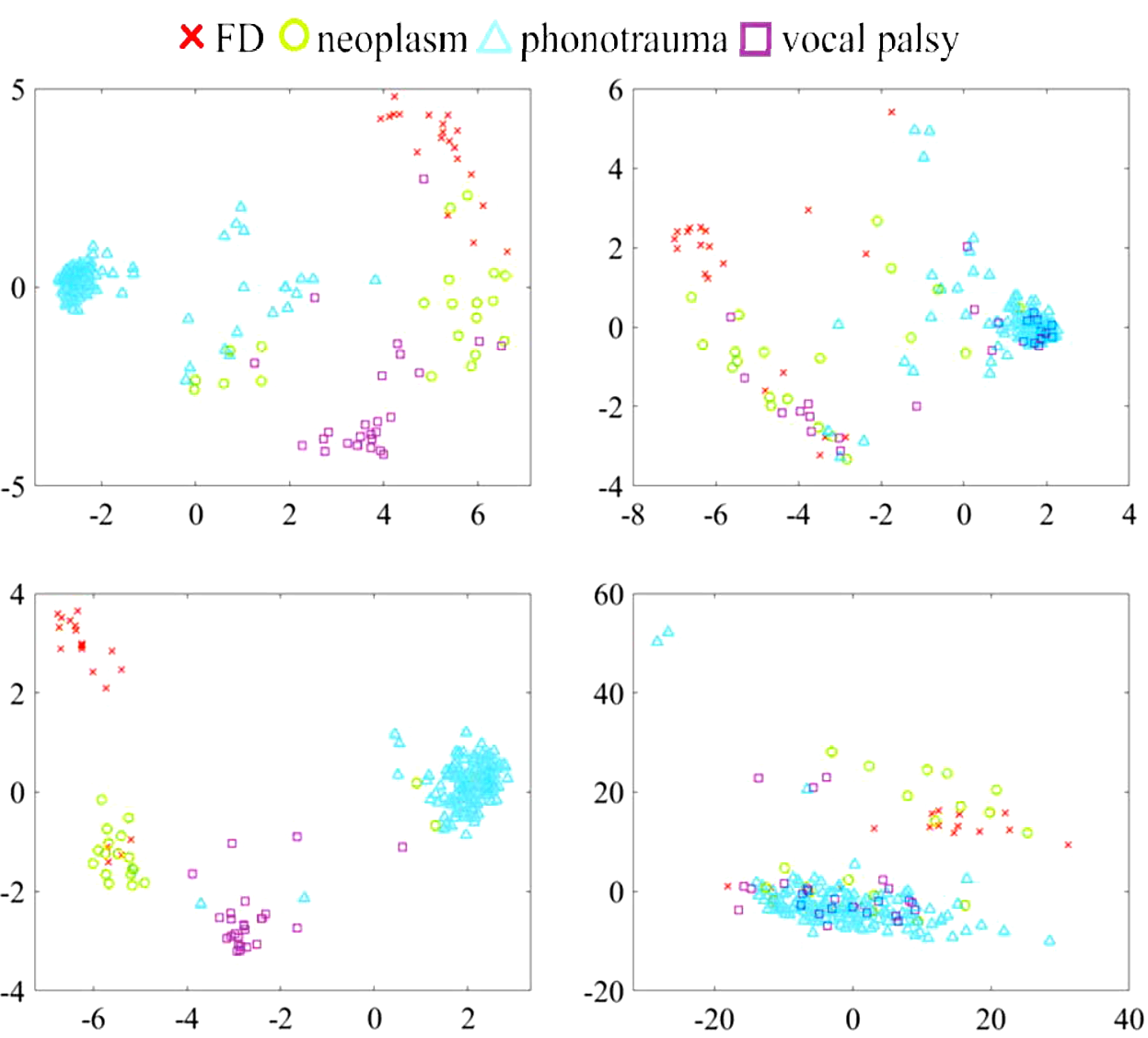}}
	\caption{The scatter plot of PCA coefficients with respect to FD, neoplasm, phonotrauma and vocal palsy clusters. From left to the right, the upper row illustrates PCA-processed $\mathbf{F}_{c,BiLSTM}$ and $\mathbf{F}_{a,BiLSTM}$, while the bottom row shows the result of those of PCA-processed $\mathbf{F}_{c,GRU}$ and $\mathbf{F}_{a,GRU}$ features.} 
	\label{fig:PCA_all}
\end{figure}

This section provides discussion from two perspectives, visualization of the disorders classification and the clinical impacts of the proposed technology. First, we carried out PCA to demonstrate system performance in the feature level from testing a model on the continuous-speech corpus. The analyses were achieved in the following steps. Each of the continuous testing utterances in FEMH was converted to MFCC streams. These frame-level MFCCs were used and passed through a classifier for extracting the acoustic features accordingly from the output of the latest hidden layer, i.e. the input of the feed-forward layer, and then averaged afterward to provide the utterance-wise representation. Finally, we collected all extracted acoustic features for PCA. For simplicity, classifier-processed utterance-level features are denoted as ``$\mathbf{F}_{c,s}$'', wherein the subscript ``$s$'' represents a classified system, that is BiLSTM or GRU in this analysis. In addition, the notation ``$c$'' denotes that this feature-extraction process was performed on a continuous database. The same procedure introduced above was also applied for extracting utterance-level features on the /$a$/-phone corpus in FEMH from each of BiLSTM and GRU and ultimately represented with the short-hand notation, ``$\mathbf{F}_{a,s}$''. Thereafter, each feature type ($\mathbf{F}_{c,BiLSTM}$, $\mathbf{F}_{c,GRU}$, $\mathbf{F}_{a,BiLSTM}$ and $\mathbf{F}_{a,GRU}$) labelled as the four vocal-disease classes, FD, neoplasm, phonotrauma and vocal palsy, were then processed by PCA for dimension reduction from 50 to 2 for further visualization. The resulting two-dimensional coefficients were depicted in Fig. \ref{fig:PCA_all}. 

From left to right, the upper row of the figure represents the PCA-processed $\mathbf{F}_{c,BiLSTM}$ and $\mathbf{F}_{a,BiLSTM}$, while the bottom two subfigures illustrated the PCA-processed $\mathbf{F}_{c,GRU}$ and $\mathbf{F}_{a,GRU}$. From the figure, we have the following observations:
\begin{itemize}
	\item The PCA coefficients of those $\mathbf{F}_{c,BiLSTM}$ and $\mathbf{F}_{c,GRU}$ features reveal more clear boundary among four classes than those of the associated $\mathbf{F}_{a,BiLSTM}$ and $\mathbf{F}_{a,GRU}$, respectively. The result shows the benefit of performing a model on continuous-speech corpus for the followed vocal-disease classifications.
	\item For all features labeled by phonotrauma, the PCA coefficients show less variety than those of other classes. Also, the small overlap between the phonotrauma cluster and others can be observed from the figure. These observations suggest the decent identified performance in the phonotrauma class.
	\item Notably, the PCA was performed on utterance-wise acoustic features and thus might not directly reflect the ultimate classification results, which was provided in terms of the frame-level majority vote from the output of classification model in Fig. \ref{fig:blockdiagram}. However, in terms of the cluster mean, we can observe the larger inter-class distance from the PCA coefficients of $\mathbf{F}_{c,BiLSTM}$ than those of $\mathbf{F}_{c,GRU}$. The observation implies that the higher sensitivity of each class from conducting BiLSTM on the continuous-speech testing condition can be obtained than those performed on the GRU system, especially for those of neoplasm and vocal palsy types.
\end{itemize}

Next, we discuss the social and clinical impacts of the proposed technology. 
Using the acoustic signal is the easiest and the most convenient way for the noninvasive screening of voice disorders.
This motivates us to use continuous speech instead of a single vowel because the multiple syllables may provide richer information to improve the performance.
Experimental results show that deep learning algorithms can detect common voice disorders using continuous speech. An alternative advantage of using continuous speech is that
it may provide valid detection information resistant to unintentionally abdominal vocalization from vocalist experts. In addition, people should minimize the contact possibility, especially during the COVID-19 pandemic period. With the proposed technology, future practice can screen patients who truly need hospital visits and reduce unnecessary medical demands, especially during the COVID-19 pandemic.

\section{Conclusion}
\textcolor{black}{This study proposes a novel pathological voice classification approach using continuous speech.}
Unlike traditional methods, which rely on the single vowel acoustic signal, continuous speech provides richer information to improve performance.
In the proposed framework, acoustic signals are transformed into MFCC features, and BiLSTM is adopted to model the sequential feature vectors. 
Experimental results demonstrated that the proposed framework yields significant accuracy and UAR improvements of 78.12–89.27\% and 50.92–80.68\%, respectively, compared with systems that use only a single vowel. 
The sensitivities for each disorder were analyzed, and the model capabilities were visualized 
via PCA. 
An alternative experiment, based on FEMH-Challenge, again confirms the advantages of using continuous speech for learning voice disorders.

\bibliographystyle{ieeetr}
\bibliography{reference}

\end{document}